\documentclass[preprints,review,accept,moreauthors,pdftex]{Definitions/mdpi}
\newcommand{\farcs}{\mbox{\ensuremath{.\!\!^{\prime\prime}}}}

\newcommand{\arcsec}{\mbox{\ensuremath{^{\prime\prime}}}}

\firstpage{1} 
\pubvolume{8}
\issuenum{6}
\articlenumber{329}
\doinum{10.3390/universe8060329}
\pubyear{2022}
\copyrightyear{2020}
\externaleditor{Academic Editor: Stephen J. Curran} 
\datereceived{2 May 2022} 
\dateaccepted{5 June 2022} 
\datepublished{14 June 2022} 
\hreflink{https://\linebreak doi.org/10.3390/universe8060329} 
\pdfoutput=1

\usepackage{comment}

\Title{The Role of Radio Observations in Studies of Infrared-Bright Galaxies: Prospects for a next-generation Very Large Array}

\TitleCitation{The Role of Radio Observations in Studies of Infrared-Bright Galaxies: Prospects for a next-generation Very Large Array}

\newcommand{\orcidauthorA}{0000-0001-7089-7325} 

\Author{Eric J. Murphy $^{1}$\orcidA{\orcidauthorA}}

\AuthorNames{Eric J. Murphy}

\AuthorCitation{Murphy, E.J.}

\address{%
$^{1}$ \quad National Radio Astronomy Observatory, 520 Edgemont Road, Charlottesville, VA 22903, USA; emurphy@nrao.edu}

\corres{Correspondence: emurphy@nrao.edu; Tel.: +1-434-296-0373}

\abstract{
The bulk of the present-day stellar mass was formed in galaxies when the universe was less than half its current age (i.e., $1 \lesssim z \lesssim 3$).   
While this likely marks one of the most critical time periods for galaxy evolution, we currently do not have a clear picture on the radial extent and distribution of cold molecular gas and associated star formation within the disks of galaxies during this epoch.  
Such observations are essential to properly estimate the efficiency at which such galaxies convert their gas into stars, as well as to account for the various energetic processes that govern this efficiency.  
Long-wavelength (i.e., far-infrared--to--radio) observations are critical to penetrate the high-levels of extinction associated with dusty, infrared-bright galaxies that are driving the stellar mass assembly at such epochs.  
In this article we discuss how the next-generation Very Large Array will take a transformative step in our understanding of galaxy formation and evolution by delivering the ability to simultaneously study the relative distributions molecular gas and star formation on sub-kpc scales unbiased by dust for large populations of typical galaxies in the early universe detected by future far-infrared space missions.  
}

\keyword{galaxies: evolution; galaxies: high-redshift; galaxies: star formation}

\begin{document}


\section{Introduction}

In standard cosmological models, baryonic material flows into the gravitational potential wells of dark matter halos as they coalesce out of an otherwise expanding universe. 
This gas will cool, condense, and eventually fuels star formation in galaxies. 
The physical processes suggested for triggering, modulating, and even suppressing star formation are numerous and still poorly understood.
On very small scales, we have observed that the instantaneously available local fuel supply must play a crucial role in star formation \citep[e.g.,][]{rck98,bigiel08,akl08}.  
Furthermore, when observed on the largest scales, the galaxy clustering environment also clearly plays an important role in regulating the rate of star formation \citep[e.g.,][]{dressler97, butcheroemler78, butcheroemler84, desai97}.  
To piece together a fully consistent, predictable theory of star formation requires tracing each of these processes on the full range of scales that they operate for a large, heterogenous set of physical conditions. 

While nearby galaxies and our own Milky Way provide the best means to study the small scale physics directly, it is critical that we relate such investigations to the intensity and distribution of star formation {\it within} large samples of "typical" 
galaxies when the majority of the present day stellar mass was produced \citep[i.e., $1 \lesssim z \lesssim 3;$][and references therein]{md14}.  
Our current inability to bridge the 
gap between star formation studied on sub-kpc scales within nearby galaxies to globally integrated measurements of large populations of galaxies at high-redshifts remains a major limitation in our understanding of galaxy formation and evolution.

Current studies that measure galaxy sizes at $z\gtrsim1$ in the rest-frame UV, optical (H$\alpha$), sub-mm (cold dust), low-frequency radio (synchrotron), and high-frequency radio (free-free) appear discrepant, suggesting different typical sizes of star-forming disks \citep[e.g.,][]{ejn16, si15, jm15, ejm17, bondi18, cotton18}.  
However, nearly all of such studies have not been carried out for the same population of galaxies, which may be a primary driver of these discrepancies.    
The lack of a reliable measurement for such a fundamental parameter such as galaxy size, let alone the distribution of star formation and cold molecular gas within them, remains a major hurdle for understanding how galaxies build up their stellar mass over cosmic time.   
Consequently, to make the transformative step into a regime of "precision" galaxy formation and evolution studies requires 
sub-arcsecond imaging of the gas distribution and dynamics, stellar mass, and star formation activity unbiased by dust for large populations of typical galaxies in the early universe.  

In this article, we provide a general summary of the next-generation Very Large Array (ngVLA), a community-defined facility that will replace both the Very Large Array (VLA) and Very Long Baseline Array (VLBA), which was recently identified as a high-priorty facility whose construction should be started this decade by the Astro2020 Decadal Survey Report \citep{astro2020}. 
We additionally identify a number of key areas where the ngVLA will make transformative steps in studies of galaxy evolution by characterizing the properties and energetics of sources likely to be detected in large far-infrared surveys.  These areas include determining: 
\begin{itemize}
    \item the size distribution of star-forming galaxy disks at $z\gtrsim1$ and how the star formation activity is distributed within them.
    \item how the distribution of star formation compares to that of the stellar mass and molecular gas at these epochs.  
    \item if the distribution of star formation is clumpy in such systems, as inferred from rest-frame UV observations of high-redshift galaxies, or is if this purely an artifact of spatially varying dust obscuration as currently indicated by observations of cold dust in such systems. 
    \item how centrally concentrated is star formation as a function environment and cosmic time.  
    \item tight constraints on the evolution of the total cold molecular gas content in typical galaxies at high (i.e., $z\gtrsim3$) redshift.  
\end{itemize}


\section{The Power of Far-Infrared Surveys}
A principal achievement of past far-infrared space missions has been the resolution of the far-infrared background into its constituent galaxies via deep surveys at mid- and far-infrared wavelengths \citep[e.g.,][]{elbaz99, papovich04, oliver10}. 
In fact, near its peak, roughly three quarters of the cosmic infrared background has been resolved into individual galaxies. 
These dusty, star-forming systems have infrared luminosities of $L_{\rm IR} > 10^{11}\,L_{\odot}$, which are often ten times larger than their optical luminosities.  
Infrared surveys from facilities including {\it IRAS} \citep{iras84}, {\it ISO} \citep{iso96}, {\it AKARI} \citep{akari07}, {\it Spitzer} \citep{spitzer04}, {\it Herschel} \citep{herschel10}, and {\it WISE} \citep{wise10} have each played a critical role in our current understanding of how such galaxies form and evolve.  
From the current epoch to $z \sim 1$, the comoving number density of these dusty star-forming galaxies increases by more than a factor of $\sim$100 \citep[e.g.,][]{ejm11a, bm13}, until they dominate the total infrared energy density at redshifts of $z\sim1-2$ when star formation activity in the universe was at its peak \citep[e.g.,][]{kc07}.  

At present, the use of far-infrared diagnostics (continuum and spectral lines) to account for both dust-obscured star formation and interstellar medium (ISM) conditions has greatly expanded from only the rare and extreme sources to more typical "main-sequence" galaxies out to $z\sim2$ \citep[e.g.,][]{de11}.  
Far-infrared diagnostics also offer key insight to disentangle star formation and accretion energetics for the hosts of active galactic nuclei (AGN). 
Consequently, far-infrared surveys of galaxies are able to characterize the various modes of star formation along with the role played by black hole growth at and since the peak of both cosmic star formation and black hole accretion activity. 

In the future, a next generation of large far-infrared imaging and spectroscopic surveys of the extragalactic sky should make it such that these powerful diagnostics can be applied to even fainter, and more numerous populations of galaxies.  
For example, recently conceived missions \citep[e.g., Galaxy Evolution Probe (GEP); {\it Origins} Space Telescope;][]{gep21, origins_sci21} aim to map the history of galaxy growth through star formation and supermassive black hole accretion, while simultaneously studying the detailed relationship between these processes \citep[e.g.,][]{pope2020wp}. 
Another key goal of such future surveys is to use the power of far-infrared fine structure lines (e.g., [O{\sc iii]}, [N{\sc iii}]) to robustly measure the growth of metals in galaxies and their evolving ISM (i.e., content and conditions) over cosmic time uninhibited by dust \citep[e.g.,][]{nagao11, jds2020wp}.  

Complementing these future large area far-infrared galaxy surveys, which will deliver volumetric demographics, with investigations of the detailed astrophysics for substantial fractions of representative sources will be a critical ingredient for constructing a physical picture for galaxy formation and evolution. 
Such detailed follow-up investigations will require sensitive, high angular resolution observations that 
can only be achieved via large radio/mm interferometers on the ground.  
Given the complementary nature of the (extinction-free) diagnostics obtained in the radio with those obtained in the infrared, joint investigations provide a powerful way to accurately distinguish and characterize star formation and black hole accretion energetics in the high-redshift universe.

\begin{figure}
    \centering
    \includegraphics[scale=0.5]{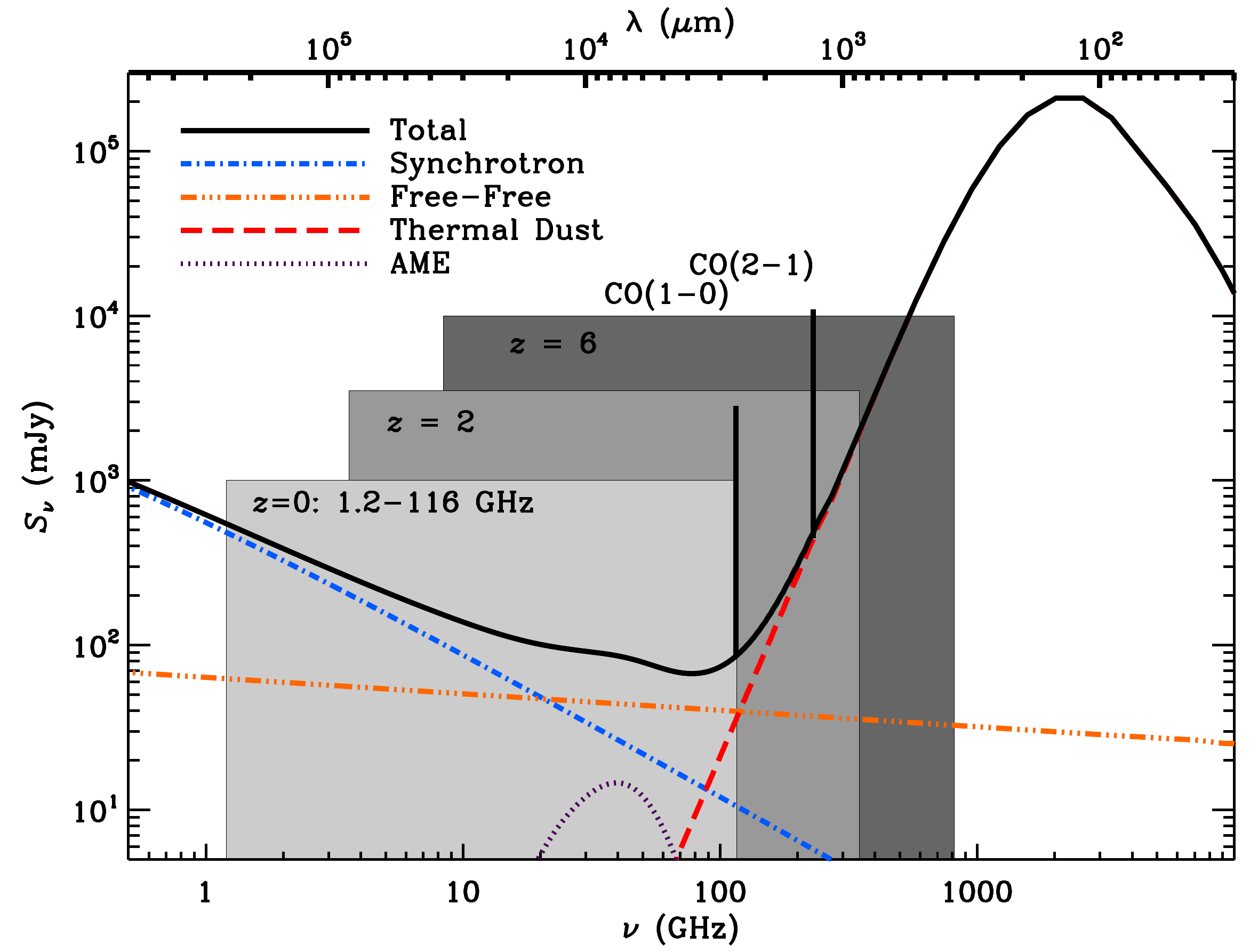}
    \caption{A model radio-to-infrared spectrum for a star-forming galaxy, showing that observations at frequencies spanning $1.2-116$\,GHz provide direct access to free-free emission, whose contribution increases with frequency, and thus with the rest-frame emission from high-z galaxies (highlighted for $z\sim2$ and $z\sim6$).}
    \label{fig:spec}
\end{figure}

\section{Radio Emission from Galaxies}

Radio continuum emission from galaxies covering $\sim1.2 - 116$\,GHz is powered by a diverse mixture of physical emission processes, each providing independent information on the star formation and ISM conditions of galaxies (see Figure \ref{fig:spec}).  
These processes include non-thermal synchrotron, free-free (thermal bremsstrahlung), thermal dust, and anomalous microwave emission (AME) that are directly related to the various phases of the ISM and provide a comprehensive picture of how galaxies convert their gas into stars.  
Each of these emission components, briefly described below, are of low surface brightness in the $\sim 30 -116$\,GHz frequency range, and therefore difficult to map in a spatially resolved manner at $\sim 10 - 1000$\,pc scales for large populations of galaxies.  
Consequently, our current knowledge about the emission processes over this frequency range is primarily limited to the brightest star-forming regions/nuclei in the most nearby sources \citep[e.g.,][]{akl11b,msc08,msc10,ejm13,ejm15,loreto15}, providing no information on how the situation may differ for ISM conditions \citep[e.g.,][]{fat18} that may be more representative of those in high-redshift galaxies, where we typically have to rely on globally integrated measurements.   

\subsection{Continuum Emission Processes at GHz Frequencies}
In the following section, we briefly describe each of the primary continuum emission components that power radio/mm emission from galaxies.  
These processes, which include non-thermal synchrotron, free-free (thermal bremsstrahlung), thermal dust, and AME, are each illustrated in Figure \ref{fig:spec}, along with the location of the $J=1\rightarrow0$ and $J=2\rightarrow1$ lines of CO. 

\subsubsection{Non-Thermal Synchrotron Emission} 
At $\sim$GHz frequencies, radio emission from galaxies is dominated by non-thermal synchrotron emission that is indirectly related to star formation. Stars more massive than $\gtrsim$8\,$M_{\odot}$ end their lives as core-collapse supernovae, whose remnants are thought to be the primary accelerators of cosmic-ray (CR) electrons, giving rise to the diffuse synchrotron emission observed from star-forming galaxies. 
Thus, the synchrotron emission observed from galaxies provides a direct probe of the relativistic (magnetic field + CRs) component of the ISM, whose role in the regulation of star formation and galaxy evolution is currently not well understood. 
As illustrated in Figure \ref{fig:spec}, the synchrotron component has a steep spectral index, typically scaling as $S_{\nu} \propto \nu^{-0.83}$ with a measured rms scatter of 0.13 \citep{nkw97}. 

\subsubsection{Free-Free Emission}  
The same massive stars whose supernovae are associated with the production of synchrotron emission in galaxy disks are also responsible for the creation of H{\sc ii} regions. 
The hot, ionized gas produces free-free emission, which is directly proportional to the production rate of ionizing (Lyman continuum) photons that is optically-thin at radio frequencies. 
In contrast to optical recombination line emission, no hard-to-estimate attenuation term is required to link the free-free emission to ionizing photon rates, making it an ideal, and perhaps the most reliable, measure of the current star formation in galaxies.  
Unlike non-thermal synchrotron emission, free-free emission has a relatively flat spectral index, scaling as $S_{\nu} \propto \nu^{-0.1}$. 

\subsubsection{Thermal Dust Emission}  
At frequencies $\gtrsim$100\,GHz, (cold) thermal dust emission on the Rayleigh-Jeans portion of the galaxy far-infrared/sub-mm spectral energy distribution can begin to take over as the dominant emission component for regions within normal star-forming galaxies. 
This in turn provides a secure handle on the cold dust content in galaxies, which dominates the total dust mass. 
For a fixed gas-to-dust ratio, the total dust mass can be used to infer a total ISM mass in galaxies \citep[e.g.,][]{heiles88,dame01,nzs16}. 

\subsubsection{Anomalous Microwave Emission}  
In addition to the standard Galactic foreground components described above, an unknown component has been found to dominate over these at microwave frequencies between $\sim 10-90$\,GHz, and is seemingly correlated with 100\,$\mu$m thermal dust emission. 
Cosmic microwave background (CMB) experiments were the first to discover the presence of AME \citep{ak96a,eml97}, whose origin still remains unknown \citep[see][for a review]{cd18}. 
Its presence as a foreground is problematic as the precise characterization and separation of foregrounds remains a major challenge for current and future CMB experiments \citep[e.g.,][]{bicep2keck2015,planck2016-XXX}. 
At present, the most widely accepted explanation for AME is the spinning dust model \citep{wce57,dl98a,planck2011-XX,hd17} in which rapidly rotating very small grains, having a nonzero electric dipole moment, produce the observed microwave emission. 

\subsection{The role for a next-generation Radio/mm Observatory}

A new radio/mm observatory covering a frequency range spanning $\sim1.2 - 116$\,GHz with an order of magnitude improvement in both sensitivity and angular resolution over existing facilities will make critical use of each of these emission processes to carry out detailed, self-consistent studies of star formation and related ISM conditions within galaxies at all epochs.  
For instance, such a facility would be sensitive to synchrotron-emitting CR electrons spanning an order of magnitude in energy (i.e., $\sim1-30$\,GeV), including the population that may drive a dynamically-important CR-pressure term in galaxies \citep[e.g.,][]{socrates08}.
Such a facility would additionally deliver robust maps highlighting the distribution of massive star formation within galaxies using optically-thin free-free emission by taking advantage of the fact that, globally, this emission component begins to dominate the total radio emission in normal star-forming galaxies at $\gtrsim$30\,GHz \citep[e.g.,][]{jc92, ejm12b}. 
By providing large instantaneous bandwidth, such a telescope would enable 
observations at $\gtrsim100$\,GHz that will simultaneously provide access to the $J=1\rightarrow0$ line of CO, revealing the molecular gas distribution and kinematics for entire galaxy disks.   
Alternatively, combining H{\sc i} observations (also obtained with by such a facility) with $J=1\rightarrow0$ CO maps, one can instead use the thermal dust emission to measure the spatially varying gas-to-dust ratio directly.  
Finally, the increased sensitivity and mapping speed of such a telescope would allow for an unprecedented investigation into the origin and prominence AME both within our own galaxy and others, ultimately helping to improve upon the precision of future CMB experiments.   

\begin{figure}
    \centering
    \includegraphics[scale=0.4]{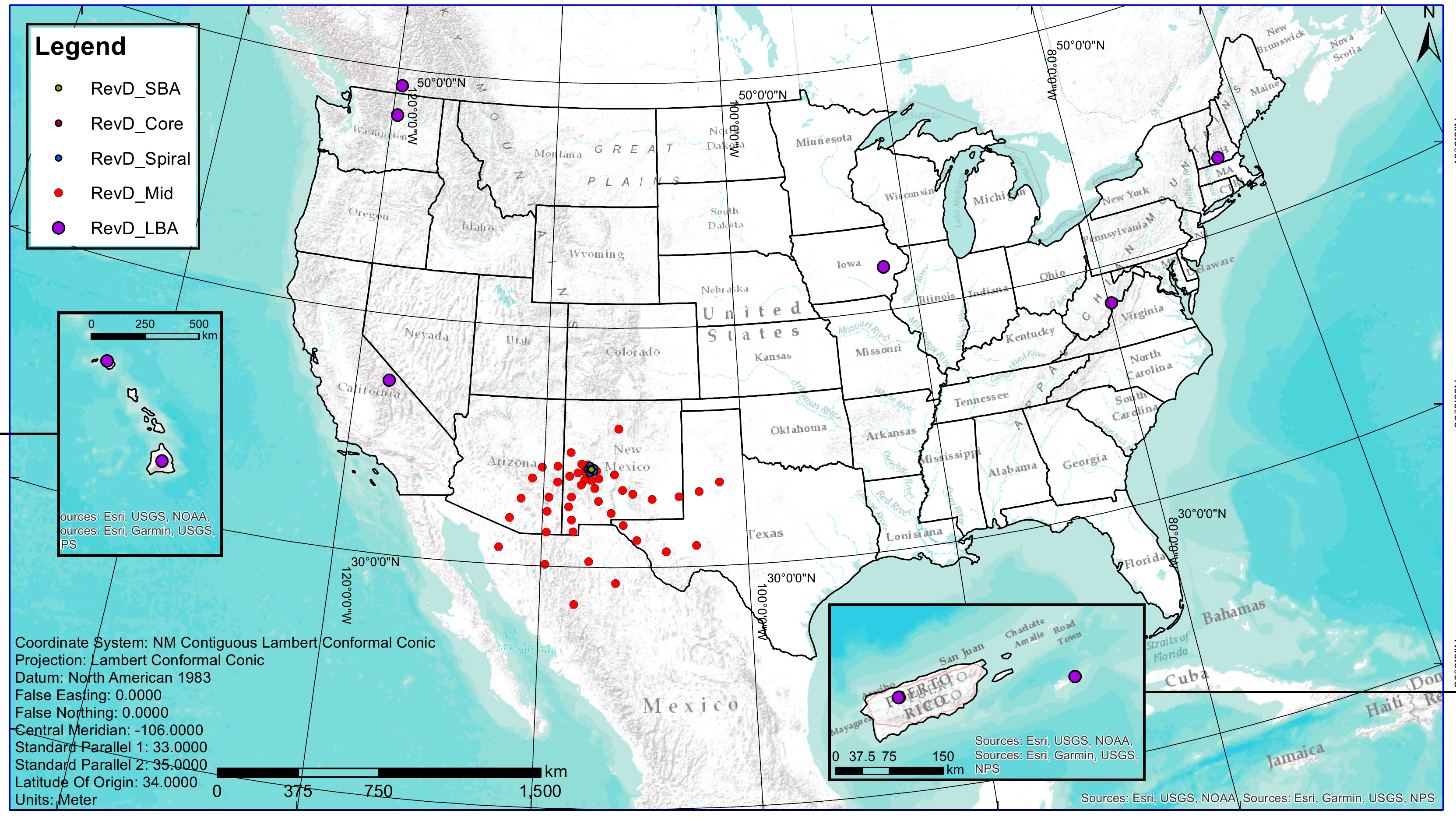}\hfill
    \includegraphics[scale=0.4]{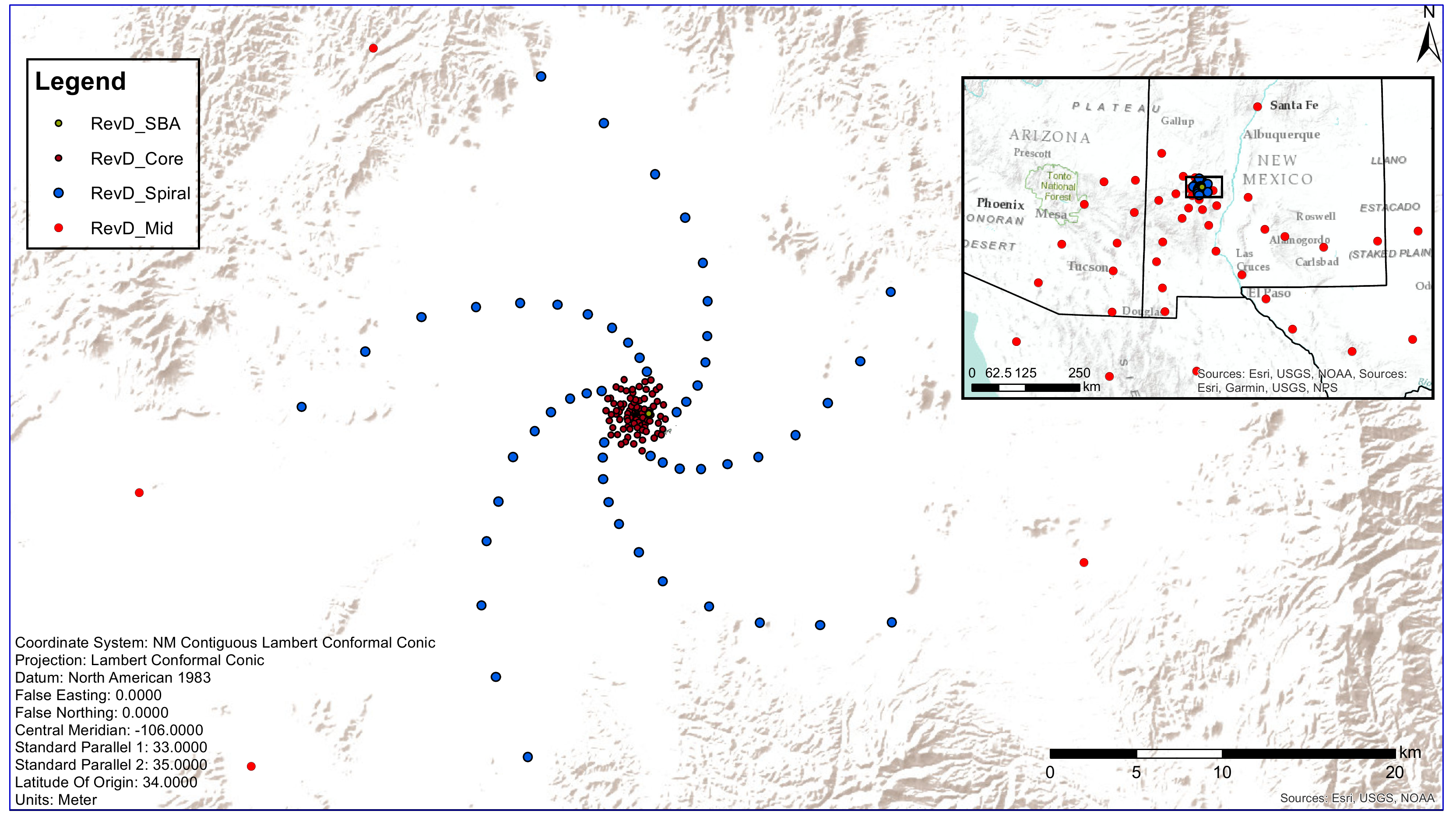}
    \caption{{\it Top:} A potential configuration layout for the ngVLA showing all 263 antennas spread across the North American Continent.  The Array is centered at the current VLA site on the plains of San Agustin in New Mexico.  The legend associates each antenna with a logistical sub-component of the full array.  Each of the LBA stations (purple dots) consists of 3 antennas.  
    {\it Bottom:} A zoom-in of the main array showing the five-are spiral pattern (54 18\,m antennas) and dense core (114 18\,m antennas).  The maximum baseline of the spiral and core antennas is 29.3 and 4.3\,km, respectively.   The 19 6\,m SBA antennas are located within the central core.  }
    \label{fig:config}
\end{figure}

\section{A next-generation Very Large Array}
To take the next major step in galaxy formation and evolution studies requires a radio/mm interferometer that is able to deliver the requisite combination of sensitivity, frequency coverage, and angular resolution to detect and resolve large populations of high-redshift galaxies unaffected by dust, as is the case for the ngVLA.  
The design of the ngVLA is the result of a close collaboration between the National Radio Astronomy Observatory and the international scientific and technical communities to define a replacement for the existing VLA and Very Long Baseline Array (VLBA).  
The ngVLA will be a transformative, multi-disciplinary scientific instrument capable of opening a new window on the universe through ultra-sensitive imaging of thermal line and continuum emission down to milliarcsecond-scale resolution, as well as unprecedented broad-band continuum polarimetric imaging of non-thermal processes (e.g., the formation and evolution of stellar and supermassive blackholes in the era of multi-messenger astronomy). The ngVLA will be optimized for observations in the spectral region between the superb performance of the Atacama Large Millimeter Array (ALMA) at sub-mm wavelengths, and the future Phase I Square Kilometer Array (SKA-1) at decimeter and longer wavelengths, thus lending itself to be highly complementary with these facilities and act as a final piece in a global suite of radio capabilities to be utilized by the entire astronomical community.  

The telescope will tackle a vast range of key, outstanding questions in modern astrophysics by simultaneously delivering the capability to: unveil the formation of Solar System analogs on terrestrial scales; probe the initial conditions for planetary systems and life with astrochemistry; chart the assembly, structure, and evolution of galaxies from the first billion years to the present; use pulsars in the Galactic Center as fundamental tests of gravity; and understand the formation and evolution of stellar and supermassive black holes in the era of multi-messenger astronomy. 
Being highly synergistic with its contemporary facilities in space, on the ground, or underground, the ngVLA will maximize the scientific returns on additional investments made by funding agencies in the U.S. and abroad.  
This is especially true for future far-infrared missions and where the joint infrared and radio capabilities can provide unique and complementary information for studies of galaxy formation and evolution.  

\subsection{Technical Description}
Building on the superb cm/mm observing conditions and existing infrastructure of the VLA site, the ngVLA has been designed as an interferometric array with ten times greater sensitivity and spatial resolution than the current VLA and ALMA, operating in the frequency range of $1.2 - 116$\,GHz.\footnote{A summary of the ngVLA performance estimates can be found here: \url{https://ngvla.nrao.edu/page/performance.}}
The full $1.2 - 116$\,GHz frequency range is covered by 6 receiver bands, housed in two separate dewars.  
The mid, low, and high frequency ranges of each receiver for the current receiver configuration is shown in Table \ref{tab:rx}.

\begin{specialtable}[H] 
\small  
\centering
\caption{The current ngVLA receiver configuration.\label{tab:rx}}
 \begin{tabular}{cccc}
    \toprule
    \textbf{Band\#}	& \textbf{$\nu_{\rm mid}$ (GHz)}	& \textbf{$\nu_{\rm high}$ (GHz)}& \textbf{$\nu_{\rm high}$ (GHz)}\\
    \midrule
    1		& 2.4 & 1.2			& 3.5\\
    2		& 8 & 3.5			& 12.3\\
    3		& 16 & 12.3			& 20.5\\
    4		& 27 & 20.5			& 34.0\\
    5		& 41 & 30.5			& 50.5\\
    6		& 93 & 70			& 116\\
    \bottomrule
    \end{tabular}
\end{specialtable}

The design of the ngVLA includes a total of 263 antennas, 244 of which are 18\,m in diameter and an additional 19 of which are 6\,m in diameter.  
In Figure \ref{fig:config} we show the currently planned ngVLA antenna configuration.  
This configuration is an update from what was originally included as part of the Reference Design submitted to the Astro2020 Decadal Survey, and improves the overall imaging performance of the telescope while still achieving all key science goals and associated science requirements.  

The configuration can naturally be divided into 3 distinct subsets based on logistical constraints such as fiber, power, and other infrastructure considerations.  
The Main Array consists of 214 18\,m antennas that are located throughout the U.S. Southwest as shown in Figure \ref{fig:config}.  
The Main Array can be further divided into: 
(i) a 4.3\,km diameter core, consisting of 114 antennas and centered at the current VLA site, 
(ii) a 5-arm spiral, consisting of 54 antennas with a maximum baseline of 39\,m (i.e., similar to the current VLA A-configuration), 
and (iii) 46 mid-baseline antennas that achieve a maximum baseline length of 1068\,km.  
The Long Baseline Array (LBA), which includes 10 sites each outfitted with 3 18\,m antennas spread across the North American Continent, achieves a maximum baseline of 8857\,km.  
Consequently, the ngVLA will greatly expand current U.S. VLBI capabilities by both replacing existing VLBA antennas/infrastructure with ngVLA technology and providing additional stations on 1000\,km baselines to bridge the gap between Main Array and present VLBA baselines.  
The Short Baseline Array (SBA) consists of 19 6\,m antennas that are used to fill in the short-spacings invisible to the Main Array.  
Note that $\approx$4 of the 18\,m antennas will be used in a Total Power mode to fill in the short baselines missed by the SBA.

\begin{figure}[h]
\centering
\includegraphics[scale=0.75]{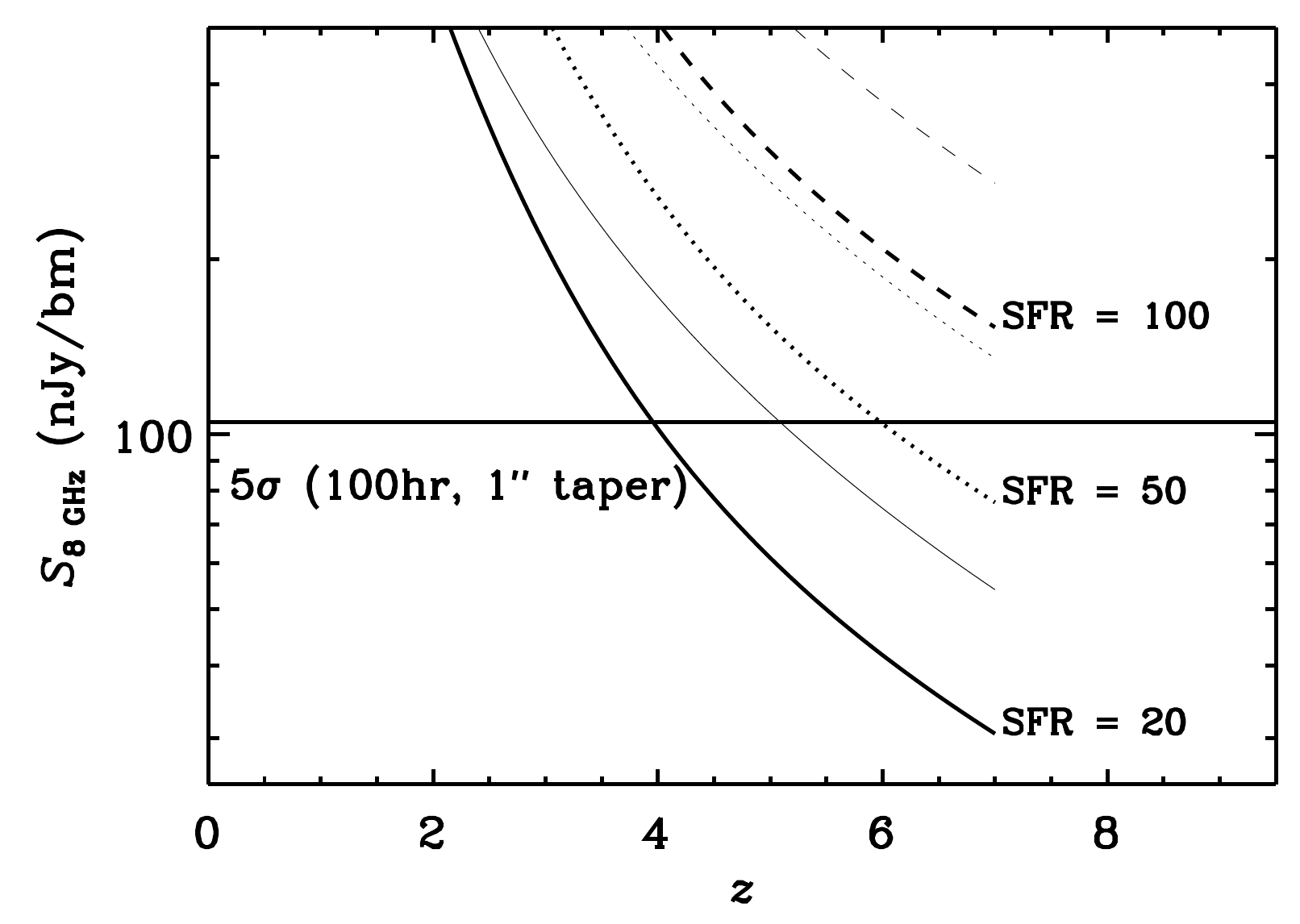}\hfill
\caption{Telescope sensitivity in nJy\,bm$^{-1}$ at 8\,GHz plotted against redshift indicating the expected brightness of a 4\,kpc disk galaxy forming stars at a rate of 20, 50, and 100\,$M_{\odot}\,\mathrm{yr}^{-1}$ with a magnetic field strength of 35\,$\mu$G \citep{ejm09c}. 
The heavy-weighted lines include estimates for synchrotron dimming due to IC scattering of CR electrons/positrons in galaxies due to the increasing CMB energy density with redshift, while the corresponding lighter-weighted lines indicate the expected brightnesses in the absence of any synchrotron dimming due to CMB effects. 
The horizontal line shows the sensitivity of the ngVLA after a 100\,hr integration and 1\arcsec~beam taper.  
The ngVLA will have enough sensitivity to detect galaxies forming stars at a rates down to $\sim 50\,M_{\odot}\,\mathrm{yr}^{-1}$ into the Epoch of Reionization, while simultaneously resolving the structure of star-forming disks throughout Cosmic Noon (i.e., $1 \lesssim z \lesssim 3$), in observations of tens to 100\,hr.   
\label{fig:sens}}
\end{figure}

\section{A New Era of Precision Galaxy Formation and Evolution Studies}
In the following section we describe a number of key areas where a next generation radio interferometer like the ngVLA will make major strides in galaxy evolution studies by providing robust measurements of the star formation activity and cold molecular gas content in galaxies over cosmic time on both global and spatially resolved scales.  

\subsection{A Tool for Robustly Measuring Star Formation at All Redshifts}
Globally, free-free emission begins to dominate the total radio emission in normal star-forming galaxies at $\gtrsim$30\,GHz \citep[e.g.,][]{jc92}, with fractional contributions as high as 80\% measured in nearby starbursts \citep[i.e., M\,82, NGC\,253, and NGC\,4945;][]{peel11}.   
Moreover, at higher redshifts, free-free emission is likely to become the dominant emission component at lower frequencies due to increased energy losses of the synchrotron emitting relativistic electron population through inverse Compton (IC) scattering off of the photons in the strengthening CMB.  
Such energy losses are linearly proportional to the energy density of the CMB, which itself increases as $\sim(1+z)^{4}$ \citep[e.g.,][]{cc01}.  
Consequently, non-thermal emission from galaxies should become severely suppressed with increasing redshift, making this frequency range ideal for accurate estimates of star formation activity in the early universe unbiased by dust \citep{jc92,ejm09c}.  

This effect is illustrated in Figure \ref{fig:sens} where telescope sensitivity in nJy\,bm$^{-1}$ at 8\,GHz is plotted against redshift indicating the expected brightness of a 4\,kpc disk galaxy forming stars at a rate of 20, 50, and 100\,$M_{\odot}\,\mathrm{yr}^{-1}$. 
The heavy-weighted lines include estimates for synchrotron dimming due to IC scattering of CR electrons/positrons in galaxies due to the increasing CMB energy density with redshift, while the corresponding lighter-weighted lines indicate the expected brightnesses in the absence of any synchrotron dimming due to CMB effects. 
This expectation of synchrotron dimming is already hinted at given recent 10\,GHz deep field observations for a sample of $z\gtrsim 1$ star-forming galaxies in GOODS-N, indicating that the rest-frame 20\,GHz flux density is dominated by free-free emission \citep[i.e., $\gtrsim 50\%$;][]{ejm17}. 
It is worth noting that there is the possibility for a contribution from AME at these frequencies \citep[e.g.,][]{ejm10, ejm18b}, which may arise from spinning dust grains \citep[e.g.,][]{dl98b}.  
However, current observations do not find a strong presence of AME both in globally-integrated and sub-kpc measurements of galaxies \citep[e.g.,][]{ejm2020wp-ame}.  

As illustrated in Figure \ref{fig:sens}, the ngVLA will have enough sensitivity to detect galaxies forming stars at a rates down to $\sim 50\,M_{\odot}\,\mathrm{yr}^{-1}$ into the Epoch of Reionization, while simultaneously resolving the structure of star-forming disks throughout Cosmic Noon (i.e., $1 \lesssim z \lesssim 3$), in observations of tens to 100\,hr.   
Accordingly, such observations will provide highly robust, extinction-free measurements of star formation rates for comparison with other optical/UV diagnostics to better understand how extinction on both galaxy-disk scales and within individual star-forming regions, evolves with redshift.  
By coupling these higher frequency observations with those at lower radio frequencies one can accurately measure radio spectral indices as a function of redshift to better characterize thermal and non-thermal energetics independently.  
This therefore enables one to determine if these physically distinct components remain in rough equilibrium with one another or if there are phsyical changes that affect the global star formation activity as a function of lookback time.  


In addition to providing information on the distribution of star formation within galaxy disks, free-free emission maps can also be used to provide detailed information on the ISM conditions for high-redshift galaxies.   
For instance, the free-free emission should prove fundamental for metallicity determinations of galaxies at all redshifts given that they provide a dust-unbiased means to normalize metal lines to the hydrogen abundance. 
As an example, combining observations of the collisionally excited pair of far-infrared [O{\sc iii}] fine-structure lines (e.g., from recently conceived far-infrared missions like GEP and {\it Origins}) with corresponding free-free emission maps may deliver the most unbiased means to measure absolute abundances in galaxies \citep[e.g.,][]{jds2020wp, lamarche22}.  
Furthermore, coupling free-free continuum maps with large, spectroscopic maps of near-/mid-infrared H-recombination lines (e.g., from {\it JWST}), which in the absence of significant dust extinction will deliver an equally robust estimate for the current star formation activity, can be used to estimate electron temperatures of the ionized gas. 
This is a critical ingredient when trying to assess metallicities from more common gas metal-abundance measures that (unlike far-infrared fine structure lines) suffer from temperature uncertainty. 
Such observations are clearly required to make the next major step for studies of star formation and chemical enrichment in the early universe.

\subsection{Measuring the Cold Molecular Gas Content of Galaxies Over Cosmic time}
\begin{figure}
    \centering
    \includegraphics[scale=0.23]{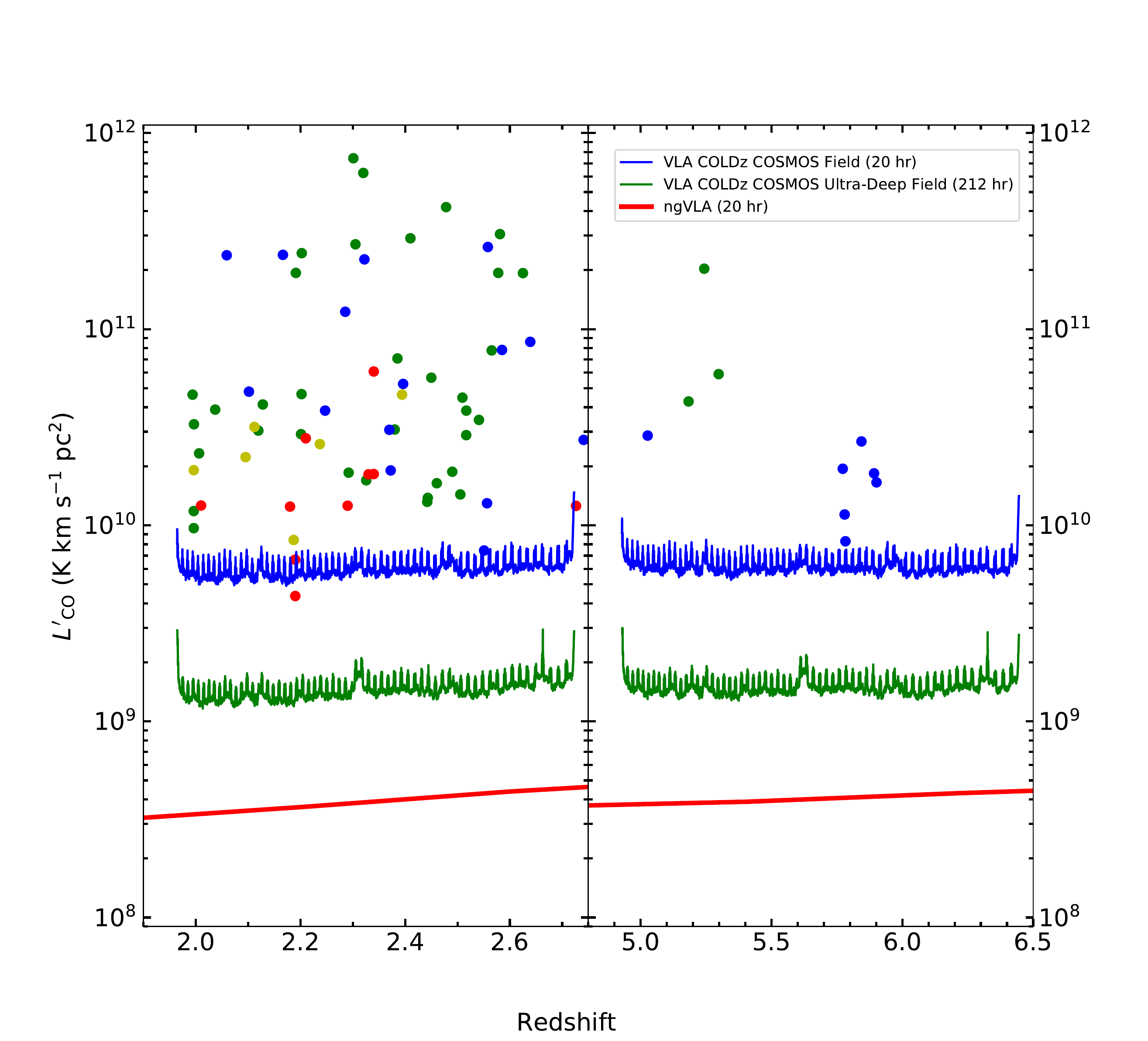}\hfill
    \includegraphics[scale=0.85]{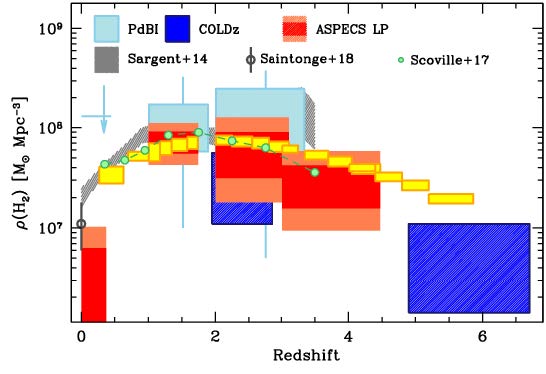}
    \caption{{\it Left --} CO line luminosity sensitivity limits (i.e., 3$\sigma$ rms noise at a line FWHM of 200\,km\,s$^{-1}$) for a number of existing surveys in the literature shown as a function of redshift. The corresponding sensitivity of the ngVLA is shown for a 20\,hr integration, which is roughly an order of magnitude better than what is currently possible.
    The different plotting colors identify different galaxy types (Green: "dusty galaxies" include sub-mm galaxies, extremely red objects, and 24\,$\mu$m selected galaxies; Red: "color-selected galaxies" include Lyman-break, BzK, and BMBX galaxies; Blue: Quasars; Yellow: Radio galaxies), which clearly indicate that significantly more sensitive observations (i.e., like those achievable with the ngVLA) are required to detect large populations of typical galaxies that represent the bulk of the luminosity function beyond $z\gtrsim3$. \citep[Adapted from][COLDz Ultra-Deep Field limit from D. Vieira et al., in preparation]{riechers19}.
    {\it Right --} The evolution of the cosmic molecular gas density showing existing constrains from NOEMA, ALMA, and the VLA that point to a similar decrease in cold gas content once beyond $z\sim2$, which is similar to what is observed for the cosmic star formation rate density over cosmic time.  
    However, current uncertainties are extremely large, demonstrating the need for a factor of $\sim$100 improvement in current survey speed to distinguish different mechanisms that can lead to the observed galaxy build-up. The ngVLA will deliver such an improvement as illustrated by the yellow data points which show the precision that would be reached by a 1000\,hr survey. \citep[Reproduced from][]{walter2020wp}}
    \label{fig:coallz}
\end{figure}

While delivering a dust-unbiased measure for the evolution of the star formation rate history of the Universe, deep radio surveys have the ability to simultaneously provide critically needed constraints on the cold molecular gas content (i.e., the fuel for star formation) over cosmic time.  
State of the art observations with existing facilities do not have the sensitivity required to detect large populations of typical star forming galaxies at high redshift (i.e., $z\gtrsim3$) that contribute to the bulk of stellar mass assembly at those epochs.  
This is shown in the left panel of Figure \ref{fig:coallz} where the CO line luminosity sensitivity limits are given for a number of existing surveys plotted as a function of redshift, highlighting various types of galaxies that have been detected (i.e., "dusty galaxies" such as sub-mm galaxies, extremely red objects, and 24\,$\mu$m selected galaxies; "color-selected galaxies" such as Lyman-break, BzK, and BMBX galaxies).  
The corresponding sensitivity of the ngVLA for a 20\,hr integration is also shown, illustrating the roughly order of magnitude improvement over what can currently be achieved.  
Beyond $z\gtrsim3$, CO has only been detected in quasar host, sub-mm, and highly magnified galaxies, providing no information on the properties for typical galaxies  that represent the bulk of the luminosity function at those redshifts.
Consequently, precise measurements of the molecular gas content of galaxies over cosmic time remain one of the last missing pieces for constructing a complete picture for galaxy formation and evolution given the already fairly tight constraints on the evolution of both the star formation and stellar mass of galaxies out to $z\sim6$ \citep[e.g.,][]{yan20}.

With the ngVLA's substantially increased sensitivity and continuous frequency coverage between $1.2-116$\,GHz, detecting typical $z\gtrsim3$ galaxies in their low-$J$ CO molecular gas will become routine and provide significantly tighter constraints on the evolution of the cold molecular gas content over cosmic time.   
This is clearly illustrated in right panel of Figure \ref{fig:coallz} where current limits on the dense gas history of the Universe are shown from a number of existing deep CO surveys \citep[i.e., ALMA ASPEcS, VLA COLDz, and NOEMA HDF-N:][]{walter14, decarli16b, riechers19, decarli20}, along with other constraints in the literature \citep{mts14, saintonge18, nzs17}.  
As with the evolution of the star formation rate density, a rise and fall is observed with redshift.  
Current observations do not provide any insight as to how the galaxies that contribute to the molecular gas density directly relate to those used to define the evolution of the cosmic density of star formation. 
Furthermore, current constraints beyond $z\gtrsim3$ are far too uncertain to accurately determine the cause for the increase of the star-formation rate at early times.

To make substantial progress in this area requires a significant increase to the amount of cosmic volume surveyed from deep spectroscopic surveys.  
For example, by achieving a factor of $\sim$100 increase in the volume that has been currently probed will in turn deliver orders of magnitude improvements of the cold gas history of the universe \citep{walter2020wp}.  
Such an improvement is illustrated in Figure \ref{fig:coallz} based on a $\sim$1000\,hr survey using the ngVLA.  
Achieving such precise measurements of the cosmic molecular gas density will put them on par with the level of precision anticipated from complementary measurements of the star formation, stellar mass, and black hole accretion histories of the universe using facilities such as $JWST$ and the US-ELTs/E-ELT.  
Such observations would also critically complement atomic line diagnostics provided by recently conceived missions far-infrared like GEP and/or an {\it Origins}-like space telescope.

\subsection{Characterizing Structure in High-Redshift Dusty Galaxies}
What is ultimately required to make the next great leap in piecing together a self-consistent theory for star formation are robust, dust-unbiased maps of the current star formation activity coupled with maps of the cold molecular gas (the fuel for star formation) for large, heterogenous samples of galaxies at $\lesssim$kpc resolution at all redshifts. 
Consequently, a key capability needed in the coming decade is high resolution ($\lesssim$0\farcs1) imaging of galaxies during the epoch of peak galaxy assembly \citep[i.e., $1\lesssim z\lesssim 3$;][]{md14}.   
It is possible that a next-generation of large far-infrared spectroscopic surveys immune to the effects of dust could be used track the co-evolution of star formation and accretion energetics over cosmic time \citep[e.g.,][]{pope2020wp}, but investigating the detailed astrophysics of individual systems will require high-resolution follow-up that is also able to penetrate high columns of dust.  
Such observations are achievable with large radio/mm interferometers on the ground, such as the ngVLA, which is specifically designed to both 
image free-free emission and low-$J$ transitions of CO from distant galaxies on these scales. 

Interferometric radio maps can achieve a range of angular resolutions required to study the size and distribution of star formation and AGN activity in populations of high redshift galaxies. 
For instance, in moderate resolution (i.e., $\sim1\arcsec$) maps, angular sizes in the radio can be used to measure star formation rate surface densities for entire galaxies that can be used to identify starbursts \citep[e.g.,][]{murray05}.  
At higher ($\sim 0\farcs01$ ) angular resolution, nuclear emission can be extracted and characterized independently, to distinguish between disk star-formation and nuclear accretion energetics \citep[e.g., through ancillary spectral diagnostics and/or brightness temperature arguments;][]{jc91}.  
This in turn provides an extremely powerful means to measure the co-evolution of star formation and black-hole accretion on a galaxy-by-galaxy basis.
Furthermore, with enough sensitivity, such high-angular resolution ($\lesssim 0\farcs01 - 0\farcs1$) observations can be used to both map the distribution of star formation on sub-kpc scales within individual high-redshift galaxies \citep[e.g.,][]{ejm2020wp-hiz}, along with the distribution and dynamics of the cold molecular gas \citep[e.g.,][]{cc2020wp}. 

\begin{figure}
    \centering
    \includegraphics[scale=0.4]{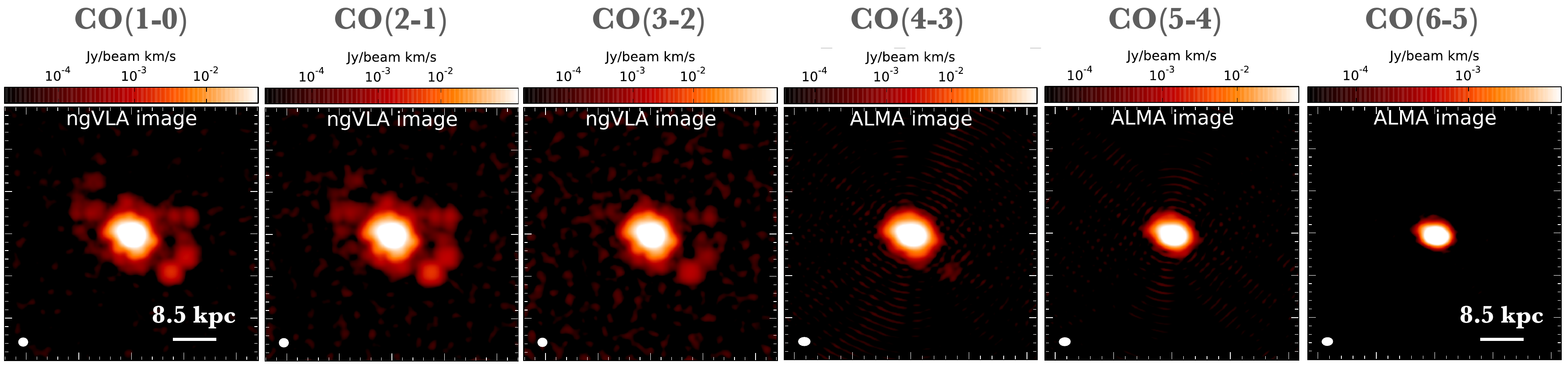}
    \caption{Simulated ngVLA and ALMA CO emission line maps for CO $J=1\rightarrow0$ to $J=6\rightarrow5$ transitions.  
    As illustrated by the simulation, access to the low-$J$ transitions by the ngVLA is critical to capture the distribution of the diffuse  emission associated with spiral arms in this $z=2$ galaxy.  
    In contrast, the ALMA-traced higher lying transitions primarily capture the warm, actively star-forming gas associated with the nuclear regions. [Adapted from D. Narayanan et al., ngVLA memo \#83]}
    \label{fig:costructure}
\end{figure}

This is illustrated in Figure \ref{fig:costructure}, where simulated ngVLA and ALMA CO emission line maps for CO $J=1\rightarrow0$ to $J=6\rightarrow5$ transitions are shown for a $z=2$ galaxy.   
What is clearly evident in Figure \ref{fig:costructure} is that by observing with the ngVLA, the majority of the CO emission from the $J=1\rightarrow0$ to $J=3\rightarrow2$ states of $z \approx 2$ galaxies are robustly recovered. 
Figure \ref{fig:costructure} additionally illustrates that the recovery of ground state emission is essential for accurate accounting of the total molecular gas inventory as galaxies become increasingly compact when observed at higher-lying transitions.  
These simulations also indicate that the highest-lying lines (e.g., $J=6\rightarrow5$) that are currently probed by ALMA may miss up to half of the underlying $H_{2}$ gas mass. 
High-resolution observations with the ngVLA will additionally be able to detect substructure, along with the ability to distinguish between physically associated clumps, and false counterparts at unrelated redshifts.  
These simulations clearly demonstrate that imaging the cool molecular gas in distant galaxies at $\lesssim1$\,kpc resolution is an essential ingredient for precision galaxy formation studies as these data provide access to the physical underpinnings of star-formation scaling relations, feedback, and the dynamics and physical conditions of the gas.  



\section{Conclusions and Future Outlooks}
The ngVLA will ultimately replace the existing VLA and VLBA, and in doing so will open a new window into studies of infrared-bright galaxies by delivering the ability to image both free-free continuum and low-$J$ CO spectral line emission from distant galaxies on sub-kpc scales at all redshifts. 
Such observations will be essential for providing deep, physical context to results from studies of large populations of galaxies detected via potential far-infrared surveys that could be possible with future missions similar to concepts such as the recently proposed Galaxy Evolution Probe and/or an {\it Origins}-like space telescope.   

Ground-based radio/mm interferometers like the ngVLA will complement such studies by providing the spatially resolved observations on sub-kpc scales that are required for a detailed understanding of how star formation is distributed in galaxy disks, how the cool molecular gas fuels that star formation, and how star formation then affects the ISM through feed-back processes. 
Such observations will complement those at other wavelengths on similar angular scales to additionally measure fundamental properties of high-redshift galaxies such as their ISM density, temperature, chemistry, and radiation fields.  
Finally, high spectral resolution observations of low-$J$ CO molecular gas will provide a new understanding on the stability of the gas disks, identify tidal features due to interactions, inflows, outflows, and derive rotation curves that can be used to probe dark-matter content of high-redshift systems.  
With such tools in hand, the next decade and beyond will truly be the beginning of a new era of precision galaxy formation and evolution studies.

\acknowledgments{I would like to thank A. Sajina and A. Cooray for inviting my participation is this Special Issue.  I would also like to thank R. Decarli, D. Riechers, and D. Vieira for providing figure material, along with the numerous co-authors of related ngVLA Science Book chapters, Astro2020 white papers, and ngVLA memos who contributed to work that is captured in this article.  
The National Radio Astronomy Observatory is a facility of the National Science Foundation operated under cooperative agreement by Associated Universities, Inc. }

\reftitle{References}


\externalbibliography{yes}
\bibliography{aph}{}


%


\end{paracol}
\end{document}